\documentclass[preprint,10pt,final,3p,times,twocolumn]{elsarticle}

\usepackage[dvipsnames]{xcolor}
\usepackage{lineno,float,caption,siunitx,amssymb,amsmath} %paralist 
\usepackage[colorlinks=true, breaklinks=true]{hyperref} % linkcolor= {green}, citecolor= {green}, urlcolor= {green}, 
\DeclareSIUnit\bar{bar}
\DeclareSIUnit\angstrom{\text {Å}}
\usepackage[singlespacing]{setspace}
\usepackage{soul} 
\modulolinenumbers[5]
\usepackage[T1]{fontenc}
\usepackage{wrapfig}
\usepackage{changes}
\journal{Nuclear Instruments and Methods in Physics Research Section A}
\begin{document}

\begin{frontmatter}



\title{The PUMA offline ion source beamline}

%% Authors

\author[TUDa]{Moritz Schlaich\corref{cor1}} %% Author name
\author[HGW]{Paul Fischer} %% Author name
\author[HGW]{Paul Florian Giesel} %% Author name
\author[TUDa,CERN]{Clara Klink} %% Author name
\author[TUDa]{Alexandre Obertelli} %% Author name
\author[HGW]{Lutz Schweikhard} %% Author name
\author[TUDa]{Frank Wienholtz} %% Author name

\cortext[cor1]{Corresponding author: \href{mailto:mschlaich@ikp.tu-darmstadt.de}{mschlaich@ikp.tu-darmstadt.de}}

\affiliation[TUDa]{organization={Technische Universität Darmstadt, Institut für Kernphysik},%Department and Organization
            addressline={Schloßgartenstraße 9}, 
            city={64289 Darmstadt},
            country={Germany}}
            
\affiliation[HGW]{organization={Universität Greifswald, Institut für Physik},%Department and Organization
            addressline={Felix-Hausdorff-Straße 6}, 
            city={17489 Greifswald},
            country={Germany}}
            
\affiliation[CERN]{organization={European Organization for Nuclear Research (CERN)},%Department and Organization
            addressline={Esplanade des Particules 1}, 
            city={1211 Geneva},
            country={Switzerland}}

%% Abstract
\begin{abstract}
The antiProton Unstable Matter Annihilation experiment (PUMA) at CERN aims to study the nucleonic composition in the matter density tail of stable and radioactive nuclei using low-energy antiprotons.
Since there is no facility in which both low-energy antiprotons and radioactive nuclei can be produced, the experimental realization with exotic nuclei requires the transportation of the antiprotons from the Extra Low ENergy Antiproton (ELENA) facility to the nearby located Isotope mass Separator On-Line DEvice (ISOLDE).
For tests and first applications of the proposed experimental technique to stable isotopes at ELENA, a dedicated offline ion source beamline was developed that will provide isotopically pure, cooled and bunched ion beams with intensities of more than $10^4$ ions per bunch while maintaining a vacuum of better than $\SI{5e-10}{\milli\bar}$ at the handover point. 
This offline ion source beamline is characterized and its capabilities are demonstrated using the example of stable krypton isotopes. 

\end{abstract}


% \begin{highlights}
% \item \textcolor{orange}{How to name PDT? energy lift or pulsed drift tube for energy modifications?}
% \item \textcolor{orange}{Do we call the sections 1,2,3... or do we say ion source section, MR-ToF MS section, RFQcb section}
% \item \textcolor{orange}{How to quantify purity needed for PUMA? --> AO}
% \end{highlights}

%% Keywords

\end{frontmatter}


%%%%%%%%%%%%%% Introduction %%%%%%%%%%%%%%
\section{Introduction}\label{sec:Introduction}
\begin{figure*}
	\centering
	\includegraphics[width = 0.96\textwidth]{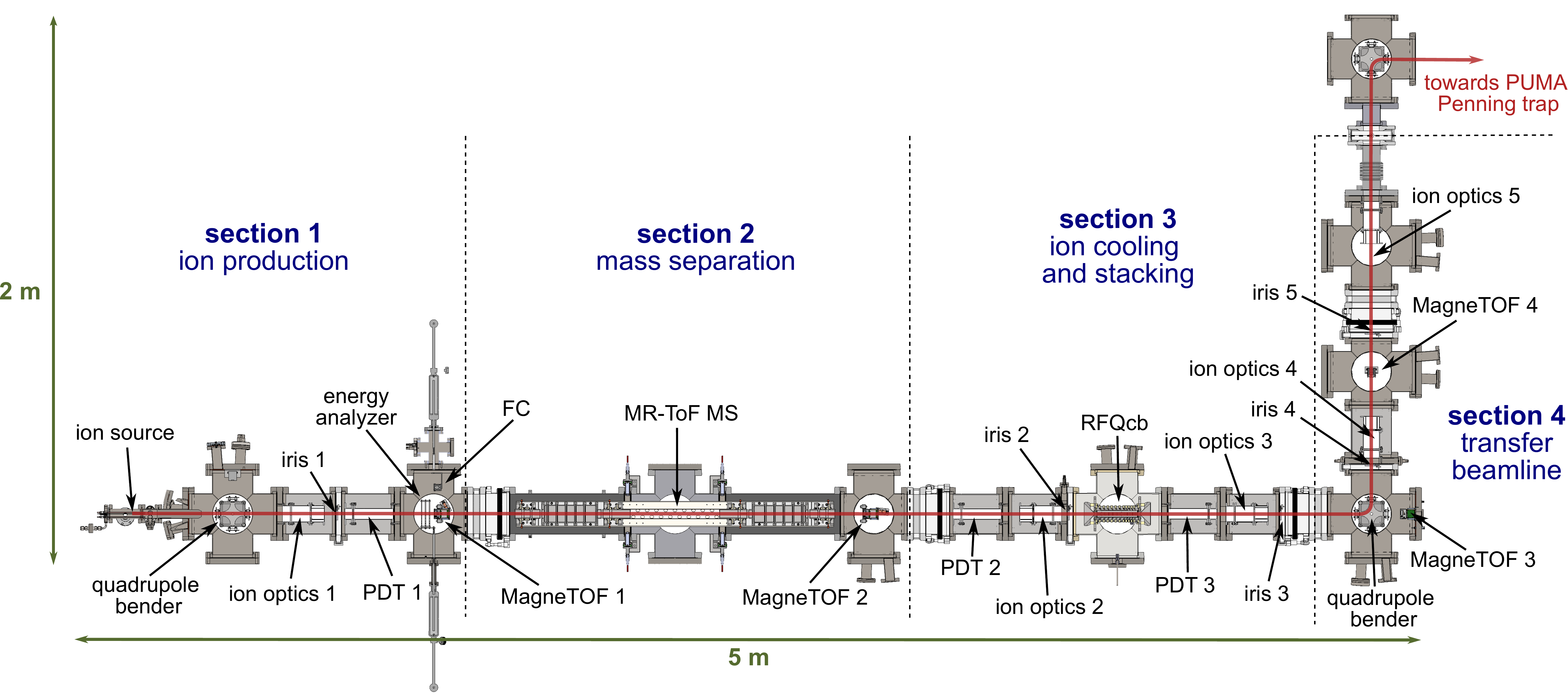}
	\caption{Top view of the PUMA offline ion source beamline CAD representation. A multi-species ion beam is generated and chopped in the ion source before it is mass-separated in the MR-ToF MS and cooled and bunched in the RFQcb. Subsequently, the purified ion bunches are guided through the transfer beamline towards the PUMA Penning trap. The four sections (indicated by dashed lines) can be separated by gate valves. Four ion detectors (MagneTOF 1-4) and a Faraday cup (FC) allow independent optimization of the ion-optical components in each section including lens and steering electrodes (ion optics 1-5) and pulsed drift tubes (PDT 1-3) for energy modifications of the ion beam. Adjustable iris apertures (iris 1-5) are used to restrict the conductance at five locations.}
	\label{fig:Offline_ion_source}
\end{figure*}
The unique and exclusive sensitivity of low-energy antiprotons interacting with the tail of the nuclear density distribution \cite{Schmidt1998NucleonRays} allows the study of phenomena that occur in the periphery of atomic nuclei such as the formation of neutron skins in neutron-rich nuclei or the occurrence of halo nuclei \cite{Tanihata1985MeasurementsRegion} close to the proton and neutron drip lines, respectively. While the understanding of neutron skins can help to constrain the nuclear equation of state \cite{Essick2021AstrophysicalAssumptions,Vinas2014DensityNuclei,RocaMaza2011NeutronExperiment}, thus connecting nuclei with large-scale nucleonic systems like, e.g., neutron stars \cite{Horowitz2001Neutron208Pb,Bertulani2019NeutronReactions}, the latter can be used to investigate nuclear few-body systems \cite{Frederico2012UniversalNuclei}. 
To this end, the PUMA experiment intends to measure the neutron-to-proton annihilation ratio on the surface of stable and radioactive nuclei via antiproton-nucleus annihilations \cite{Aumann2022PUMACollaboration}. \\
\indent
The experimental realization of the proposed technique requires the formation of antiprotonic atoms \cite{Wada2004TechnicalNuclei,Bugg1973EvidenceAbsorption} with the nuclei of interest, which is achieved by trapping both the respective ions and the antiprotons together in a nested Penning trap \cite{GABRIELSE1988nestedPenning,Hall1996nestedPenning}. Following the antiproton capture in a high-energy orbital of the atom, the antiproton decays into lower energy states until it reaches the nuclear surface. 
As a result, the antiproton annihilates with either a proton or neutron located in the tail of the nuclear density distribution. 
The study of the annihilation products provides the neutron-to-proton annihilation ratio as an observable for the characterization of the isospin asymmetry in the density tail of atomic nuclei.\\
\indent
The PUMA experiment will be conducted at the European Organization for Nuclear Research (CERN) where the low-energy antiprotons will be obtained from the Extra Low ENergy Antiproton (ELENA) facility \cite{Maury2014ELENA:CERN,Carli2022ELENA:Physics} at the Antimatter Factory while the radioactive ions will be provided by the Isotope mass Separator On-Line DEvice (ISOLDE) \cite{Catherall2017TheFacility}. 
The PUMA Penning trap is integrated in a transportable frame so that antiprotons can be accumulated at ELENA and brought to ISOLDE, where the experiment with radioactive nuclei is performed. 
For an application of the proposed technique independent of the antiproton transport, the measurements at ISOLDE are complemented by an experimental program on stable nuclei at ELENA. 
Consequently, the PUMA experiment requires a versatile off\-line ion source at ELENA that is capable of producing a broad range of ion species where each is provided with isotopic purity to allow isotope-selective measurements. 
As the rate of annihilation with the isotopes of interest must exceed the rate of annihilations with residual gas particles, which should be less than $\SI{1}{\hertz}$ inside the Penning trap \cite{Aumann2022PUMACollaboration}, the injection of the ions with an intensity of more than $10^4$ ions per bunch is required. 
Furthermore, the ion source must be operated at a pressure below $\SI{5e-10}{\milli\bar}$, measured at the junction to the antiproton beamline, to meet the strict vacuum requirements. 
This value has been determined by simulations for the condition that the pressure in front of the PUMA trap stays below $\SI{e-11}{\milli\bar}$, which would ensure antiproton storage times of more than 100 days. \cite{Aumann2022PUMACollaboration}.

%%%%%%%%%%%%%% POLIS experiment overview %%%%%%%%%%%%%%
\section{Experimental setup}\label{sec:OIS_overview}

The PUMA OffLine Ion Source (POLIS) beamline consists of four main sections. An overview of the experimental setup is shown in Fig. \ref{fig:Offline_ion_source}. 
In the first section, an electron-impact ionization source (SPECS Surface Nano Analysis GmbH, IQE 12/38) produces ions from gaseous target materials. 
During typical operation, a continuous ion beam is extracted from the ionization region at $\SI{3}{\kilo\electronvolt}$, which is chopped by a beam deflector (see Sec. \ref{sec:ionsource}).
It is followed by a quadrupole ion bender that will enable the addition of further ion sources perpendicular to the beamline. 
It is foreseen to equip the section with a surface ionization source and a laser ablation ion source at a later stage. 
\begin{figure*}[ht]
	\centering
	\includegraphics[width = 0.96\textwidth]{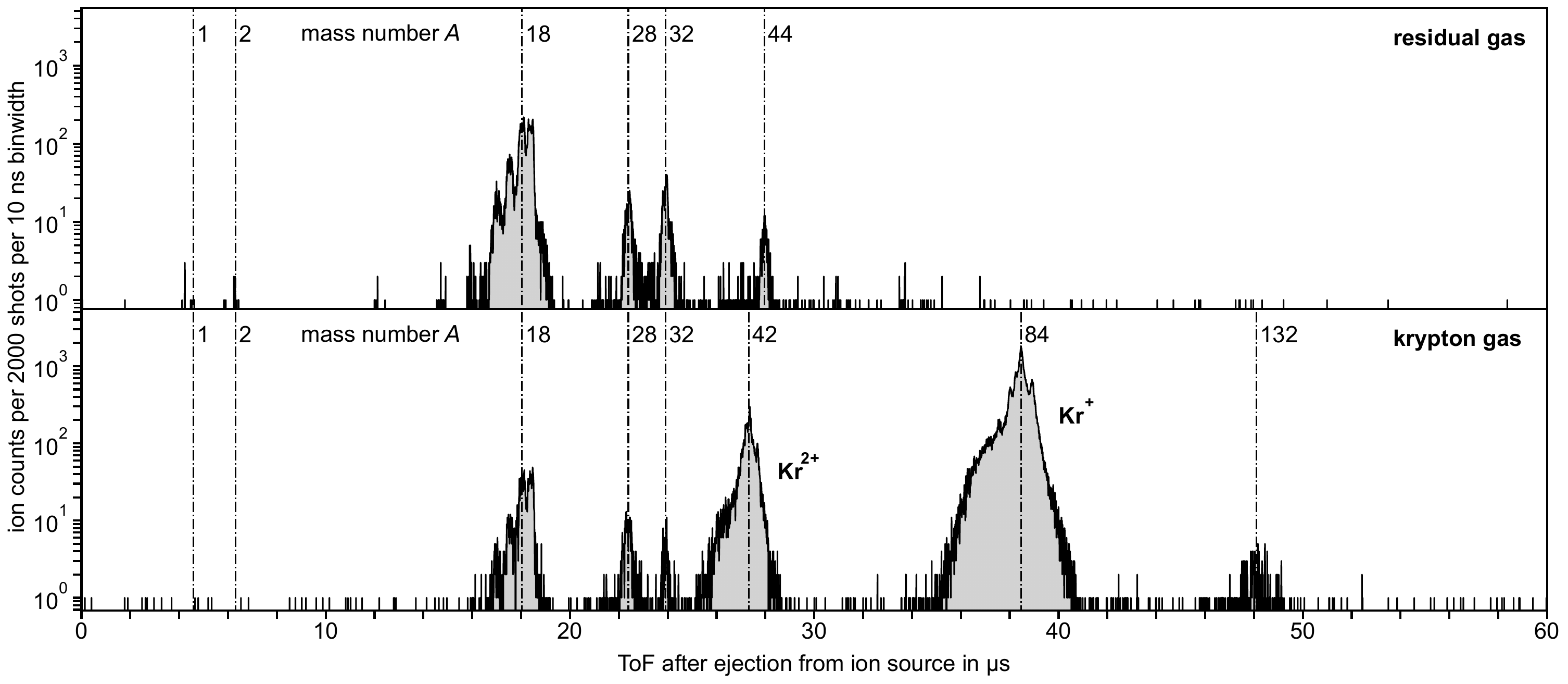}
	\caption{Time-of-flight spectra of chopped ions extracted from the ion source using the residual gas in the ionization chamber (top) and krypton gas (bottom) as target material. Guiding the ions through the MR-ToF MS without being captured, the signal is measured using MagneTOF 2. See text for the ion species corresponding to the indicated mass numbers and further details.}
	\label{fig:shoot_through}
\end{figure*}
\\
\indent
Downstream of the ion bender, ion optical elements guide the ion bunches into a multi-reflection time-of-flight mass spectrometer (MR-ToF MS) \cite{Schlaich2024AExperiment}, where they are captured using the in-trap lift technique \cite{Wolf2012Static-mirrorLift} and separated according to their mass-over-charge ratio. In-trap deflector electrodes synchronized with the revolution frequency of the ions of interest deflect unwanted ion species, ensuring the transmission of the ions of interest only \cite{Fischer2018In-depthDevice}.\\
\indent
Once ejected from the MR-ToF MS, again using the in-trap lift technique \cite{Wienholtz2017Mass-selectiveLift}, the purified ions traverse another set of ion optical components, where they are decelerated to about $\SI{300}{\electronvolt}$ using a pulsed drift tube (PDT) and injected into a radio-frequency quadrupole cooler-buncher (RFQcb). Repeating this sequence, multiple purified ion packets can thus be accumulated in the RFQcb until the targeted ion intensity of >$10^4$ ions per bunch is reached. 
This has the advantage that the initial intensity of the continuous ion beam can be reduced, which has a positive effect on the lifetime of the ion source.
Efficient ion accumulation requires the introduction of a buffer gas, in the present study helium, nitrogen or argon. Collisions between the ions and the neutral buffer gas atoms cool the ions into the minimum of the axial potential profile created by the RFQcb DC electrodes (see Sec. \ref{sec:RFQcb}) and thus reduce the longitudinal and transverse emittance of the ion bunch. \\
\indent
After the accumulation phase, the ion bunch is released from the RFQcb and reaccelerated to an energy of up to $\SI{5}{\kilo\electronvolt}$, optimized with respect to the transmission into the PUMA Penning trap. 
After the reacceleration, ions are transmitted through two $\SI{90}{\degree}$ quadrupole ion benders. The section in between, which is below referred to as the transfer beamline, is intended to gradually lower the vacuum and reach the required pressure of $\SI{5e-10}{\milli\bar}$ at the junction to the antiproton beamline. Besides the two $\SI{90}{\degree}$ bends that reduce the propagation of the RFQcb buffer gas, adjustable iris apertures define three differentially pumped vacuum sections, each of which is pumped with a turbomolecular pump (Agilent Technologies, Inc., TwisTorr 704 FS, 660 l/s for N$_2$ and TwisTorr 305 FS, 250 l/s for N$_2$), while the last section is pumped with an ion getter pump (Agilent Technologies, Inc., Vaclon Plus 300 StarCell, 240 l/s for N$_2$).

%%%%%%%%%%%%%% Standard ion optical components %%%%%%%%%%%%%%
\subsection{Basic beamline components}\label{sec:standard_components}
Throughout the beamline, different ion-optical components are used to manipulate the beam or the vacuum conditions. 
These are five electrode assemblies, each combining an einzel lens with a fourfold segmented electrode for additional steering, three pulsed drift tubes for ion energy modifications, and five adjustable iris apertures to restrict the gas flow between the individual sections. 
A detailed characterization and description of the components can be found in \cite{Klink2024DevelopmentKeV}.

%%%%%%%%%%%%%% Ion detectors and data acquisition %%%%%%%%%%%%%%
\subsection{Ion detectors \& data acquisition}\label{sec:detectors}
For an independent characterization of the four POLIS sections, it is essential to have separate ion beam diagnostics. Therefore, four MagneTOF ion detectors \cite{Stresau2006ARange} (ETP Electron Multiplier Pty Ltd, models 14924, 14925 and 14DM584) are used for time-of-flight (ToF) measurements along the ions' path (see Fig. \ref{fig:Offline_ion_source}). These detectors use magnetic and electric fields to guide electrons created by the impact of incoming ions through a cascade of multiplication stages. The gain of the electron multiplication depends on the operation voltage $U_\text{d}$, which in the case of the POLIS setup is typically between $\SI{-2.2}{\kilo\volt}$ and $\SI{-2.5}{\kilo\volt}$ (for unaged detectors). The detectors can be used for single-ion counting, but also for recording their analog signals to estimate the number of ions in high-intensity bunches, which is essential for the POLIS beamline. \\
\indent
Similar to the comprehensive MagneTOF characterization in \cite{Simke2024EvaluationBunches}, the raw signal is directly recorded by an oscilloscope (Teledyne LeCroy, WaveRunner 9104, sample rate: $\SI{20}{\giga\text{S}\per\second}$, bandwidth: $\SI{4}{\giga\hertz}$) and further analyzed with a Python-based evaluation program. This allows the application of two different analysis methods. If the ion intensity per bunch is low, i.e. below a few tens of ions in a bunch with a width of typically $\approx\SI{300}{\nano\second}$, the ToF information of the single-ion impacts is saved so that ToF measurements can be performed with sub-nanosecond resolution. In case of higher intensities, the probability for overlapping of the individual ion signals increases significantly, which precludes a precise determination of the number of ions by single-ion counting. Instead, the detector signal is integrated over time. For a given $U_\text{d}$, the number of ions $N_\text{ion}$ is proportional to the integral of the detector signal $A_\text{det}$,
\begin{equation}\label{eq:calibration}
    N_\text{ion} = c_\text{cal}\cdot A_\text{det}.
\end{equation}
The calibration factor $c_\text{cal}$ depends on $U_\text{d}$ and is determined by fitting the data of a low-intensity measurement (see \cite{Simke2024EvaluationBunches}).

%%%%%%%%%%%%%% Ion Source Characterization %%%%%%%%%%%%%%
\subsection{Electron-impact ionization source }\label{sec:ionsource}
The electron-impact ionization source generates ions from gaseous targets.~%
The neutral gas atoms or molecules injected continuously into the ionization chamber are ionized by the impact of electrons which are released from a ring-shaped cathode and accelerated to $\SI{100}{\electronvolt}$. 
Positive ions are formed and subsequently extracted by a user-defined potential between $\SI{0.2}{\kilo\electronvolt}$ and $\SI{5}{\kilo\electronvolt}$. 
Depending on the electron emission current and target gas pressure, a continuous ion beam of a few pA to a few \textmu A is extracted, at pressure levels between <$\SI{e-9}{\milli\bar}$ and $\SI{e-8}{\milli\bar}$ as measured in the six-way cross at which the source assembly is mounted. 
These pressure values are mainly influenced by the flow rate of the gas target and are only that low due to differential pumping of the ionization chamber. 
Using a solid-state switch (Behlke Power Electronics GmbH, HTS-11-07-HB-C) operated at a rate of $\sim\SI{10}{\hertz}$, the continuous beam is chopped into ion packets of a few hundred nanoseconds full width at half maximum (FWHM) by fast switching between a deflection and transmission state of deflector electrodes that are included in the electron-impact ionization source \cite{SPECSSurfaceNanoAnalysisGmbH2013IQEV2.2}. \\
\indent
In Fig. \ref{fig:shoot_through}, a typical ToF spectrum of the chopped ion beam is shown. In the presented case, ions are transmitted through the MR-ToF MS (single-path mode) and are detected using MagneTOF 2 (see Fig. \ref{fig:Offline_ion_source}). 
Target-gas ions with charge state $z=1$ and $z=2$ cover the majority of the observed signal. 
However, other ion species appear in the spectrum that originate from residual gas ionizations either directly by electron impact or by charge transfer from a primary electron-impact ion.
Besides ions with $A/z=28$, assigned to both N$_2^+$ and CO$^+$, and with $A/z=32$, assigned to O$_2^+$, several ion species are found around $A\simeq18$ (indicative of water molecules, including fragments and protonated species). 
With less abundance, ions with $A/z=44$, assigned to CO$_2^+$ \cite{BENNETT2004149}, as well as traces of protons and H$_2^+$, are produced from the residual gas.
The signals around $A/z=132$ are probably due to xenon, as this gas has been used prior to the krypton measurements.

\begin{figure}[ht!]
	\centering
	\includegraphics[width = 0.48\textwidth]{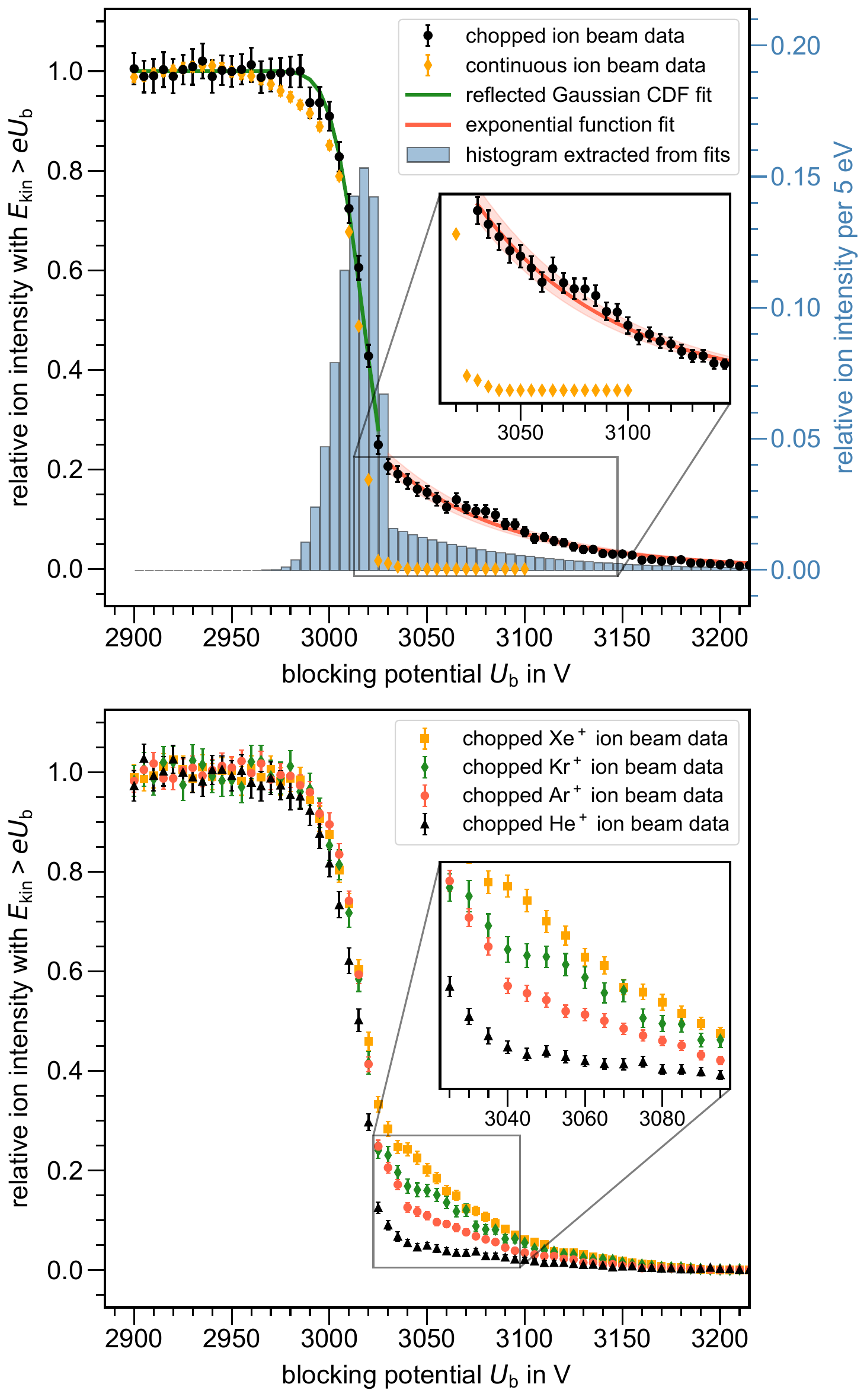}
	\caption{Relative ion beam intensity transmitted through the energy-analyzer grids as a function of the blocking potential applied to the center grid. For the example of a chopped Kr$^+$ beam (top), the combination of a reflected Gaussian cumulative distribution function (CDF; green), and an exponential function (red) has been fitted to the experimental data (black circles). A histogram extracted from the fits visualizes the energy distribution of the beam. For comparison, the result of a measurement with a continuous Kr$^+$ beam (yellow diamonds) is added to the figure. Further comparing the measurement for ion species of different masses, the results of chopped He$^+$, Ar$^+$, Kr$^+$ and Xe$^+$ beams are plotted (bottom).}
    \label{fig:energy distribution}
\end{figure}

Besides the mean kinetic energy, it is valuable to know the width and shape of the energy distribution of the extracted ion beam as it has a significant influence on the performance of the MR-ToF MS and the trapping efficiency of the RFQcb. 
These ion-energy properties can be estimated using the energy analyzer located in the first section, which has already been employed for the energy-distribution determination of the antiprotons obtained from ELENA \cite{Fischer2024DesignExperiment}. 
The energy distribution is determined by measuring the ion transmission as a function of the blocking potential created by the energy analyzer.  
This is visualized in Fig. \ref{fig:energy distribution} for continuous and chopped-beam operation using the example of Kr$^+$ ions (top) and comparing the results obtained from different ion species across the available mass range (bottom). 
For all measurements, the kinetic energy was set to $\SI{3}{\kilo\electronvolt}$ and the chopped ion packets were created with a transmission pulse of $\SI{250}{\nano\second}$ applied to the ion-source deflector electrode. 
In the case of chopped-beam operation, a high-energy tail is noticeable, as also visualized by the histogram of the ions' energy distribution (see top of Fig. \ref{fig:energy distribution}), which is almost completely suppressed in the case of continuous beam operation. 
It suggests that the switching of the deflector electrodes imprints additional energy to a relevant fraction of the transmitted ions.
This characteristic feature is visible for all ion species tested, with the relative intensity in the tail decreasing with the ion mass.
The Gaussian cumulative distribution function (CDF) used to fit the energy distribution outside the high-energy tail leads to a mean value of $\SI{3017.5(3)}{\electronvolt}$ and a standard deviation of $\SI{12.9(4)}{\electronvolt}$ for Kr$^+$ ions.

%%%%%%%%%%%%%% MR-ToF MS Characterization %%%%%%%%%%%%%%
\subsection{Multi-reflection time-of-flight mass spectrometer}\label{sec:MR_ToF}
\begin{figure}[ht]
	\centering
	\includegraphics[width = 0.48\textwidth]{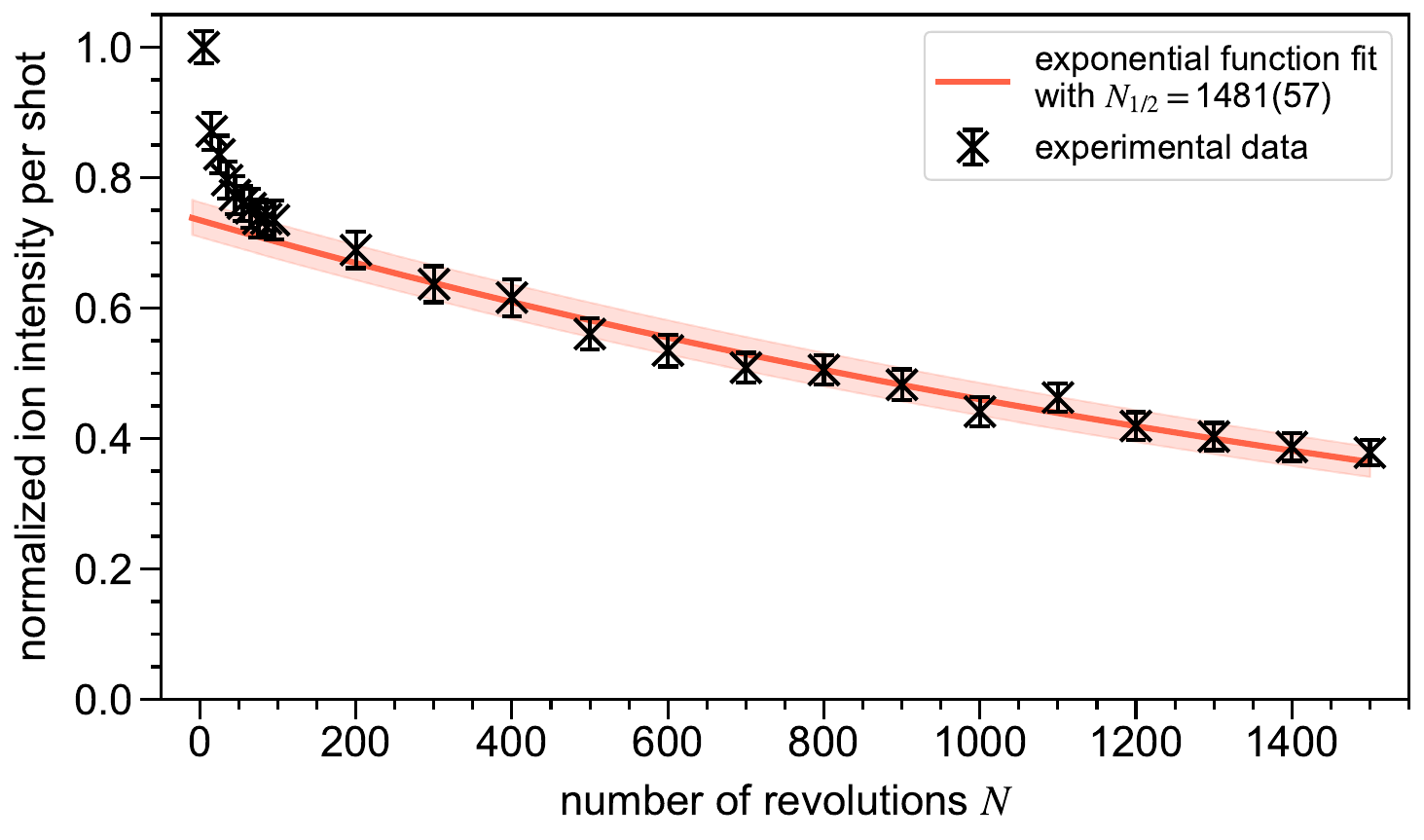}
	\caption{Normalized intensity of the ion beam extracted from the MR-ToF MS as function of the number of revolutions $N$. An exponential model is used to fit the experimental data for $N\geq 150$.}
	\label{fig:MR-ToF_transmission}
\end{figure}

The multi-species ion beam ejected from the source is to be purified in the MR-ToF MS, which is connected downstream of the ion-source section. 
It provides mass separation of any ion of interest injected with isotopic purity.~%
A detailed discussion of the PUMA MR-ToF MS development and first characterizing measurements have already been given in \cite{Schlaich2024AExperiment}. 
\begin{figure}[ht]
	\centering
	\includegraphics[width = 0.48\textwidth]{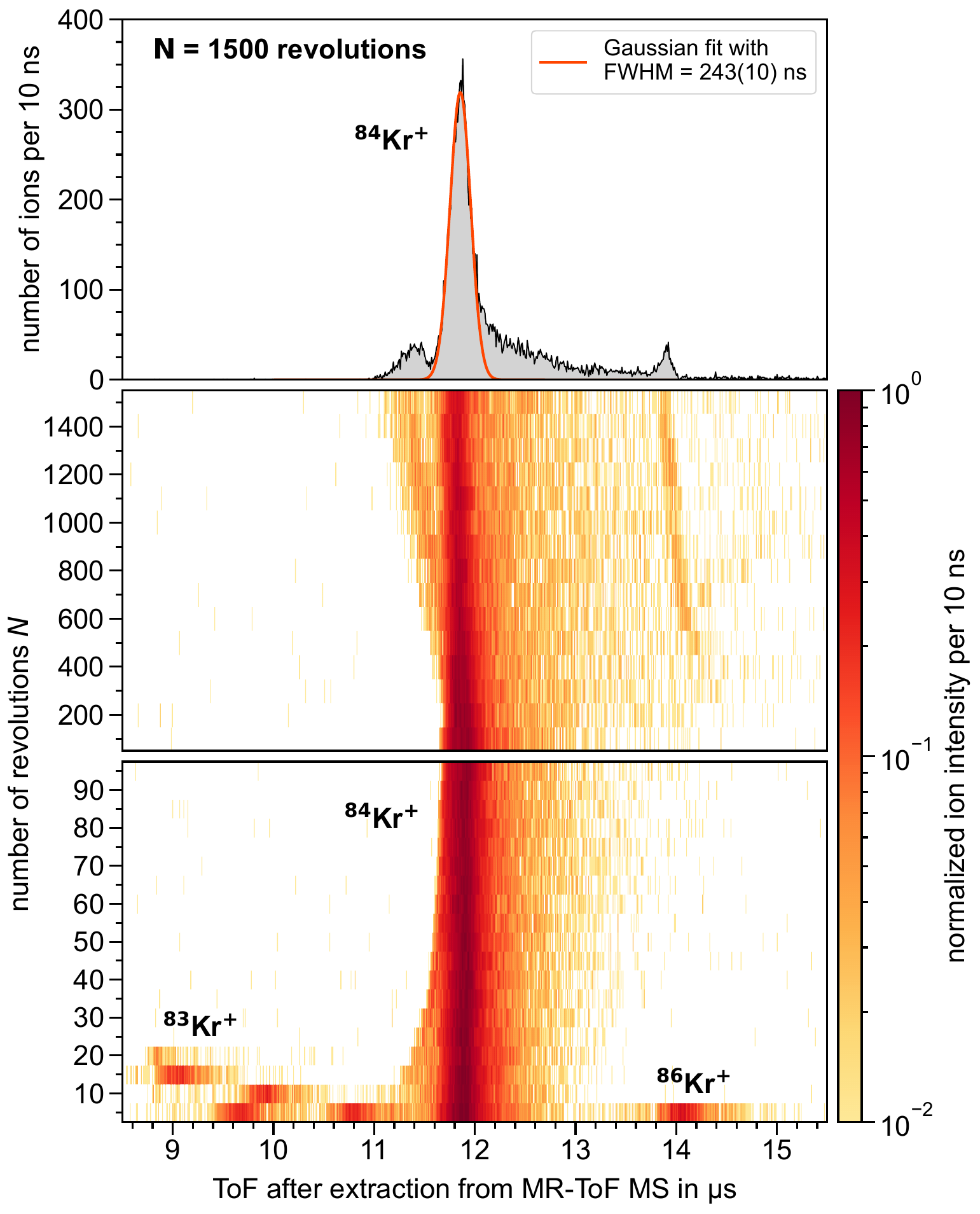}
	\caption{Ion intensity of $^{84}$Kr$^+$ ions normalized to the maximum count rate per $\SI{10}{\nano\second}$ as a function of the MR-ToF MS revolution number ($N$) and the ToF after their extraction from the MR-ToF MS. For $N<20$, the krypton isotopes $^{82,83,86}$Kr$^+$ can be seen, for $N>20$ they are removed by the deflector (bottom). For larger revolution numbers starting from $N>300$, a fraction of the beam cannot be focused sufficiently leading to a broadening of the signal (center), while the width of the main signal can be maintained up to $N=1500$ revolutions (top). For each measurement, the in-trap lift ejection and the in-trap deflector were synchronized to the $^{84}$Kr$^+$ revolution period of $T=\SI{29.0332}{\micro\second}$.}
	\label{fig:N-vs-ToF}
\end{figure}
\\
\indent
Typically, the isotopic purification process requires only a few tens of revolutions, which is equivalent to a flight time of approximately $\SI{1}{\milli\second}$ in the case of the heaviest ions tested (Xe$^+$) or less for lighter ion species. 
However, to reach higher mass resolving powers that allow isobaric separation, the ions need to be reflected for up to several thousand revolutions \cite{Wienholtz2020ImprovedStabilization,Nies2025mass103Sn,Nies2023mass99In}. 
This requires excellent stability of the individual mirror electrode potentials that are precisely chosen to compensate for kinetic energy differences that originate from the source-specific energy distribution. 
In the present case, the beam chopping has to be considered too (see Sec. \ref{sec:ionsource}). To passively stabilize the electrostatic mirror potentials, low-pass filters were added to all positively biased mirror electrodes \cite{Wienholtz2020ImprovedStabilization}. Active potential stabilization \cite{Wienholtz2020ImprovedStabilization,Fischer2021MultipleSpectrometry} can be added in the future, if the need arises, but was not necessary for the present studies. \\
\indent
As an example of the MR-ToF MS performance, Fig. \ref{fig:N-vs-ToF} shows data of a 3-$\SI{}{\kilo\electronvolt}$ Kr$^+$ beam chopped with a 300-$\SI{}{\nano\second}$ gate. 
Color-coded ToF spectra measured with MagneTOF 2, are plotted as a function of the number of revolutions $N$ of the $^{84}$Kr$^{+}$ stored in the MR-ToF MS (center and bottom). 
Additionally, the ToF spectrum for $N=1500$ is shown (top) to demonstrate the typically achievable mass resolving power and thus separation capability of the current system (see below).
\\
\indent
Besides the ion of interest ($^{84}$Kr$^+$), contaminant krypton isotopes ($^{82,83,86}$Kr$^+$) can be seen in the first 20 revolutions before they are removed by the deflector. 
For longer storage times, starting from about 25 to 30 revolutions, $^{84}$Kr$^+$ is isolated completely and thus prepared to be transmitted to the RFQcb for accumulation. 
With increasing $N$, a pronounced tailing to the right side of the signal builds up. 
It is understood by ions located in the high-energy tail (compare Sec. \ref{sec:ionsource}), which, for the potential profile used, penetrate deeper into the electrostatic mirror and therefore require more time for a revolution. 
Starting at about $N=500$, these tail ions are affected by the switching edges of the deflector, which gives rise to a peak-like artifact about $\SI{2}{\micro\second}$ after the $^{84}$Kr$^+$ signal.
Nevertheless, even at $N=1500$, the main part of the beam can be kept at the initial signal width of about $\SI{250}{\nano\second}$ corresponding to a mass resolving power of almost $10^5$.\\
\indent
In Fig. \ref{fig:MR-ToF_transmission}, the transmission is plotted as a function of the revolution number to monitor the ion loss during the purification process. 
The evaluation is based on the data shown in Fig. \ref{fig:N-vs-ToF}, where the counts for $N\leq 20$ corresponding to ions other than $^{84}$Kr$^+$ were excluded based on their natural abundance ratio.
It has been normalized to the first data point, which belongs to a storage for five revolutions. 
From the plot, two phases can be distinguished: 
Within the first 100 revolutions, the ion intensity drops by about $\SI{25}{\percent}$, which is interpreted as the loss of ions that do not fit into the acceptance phase space of the MR-ToF MS.
After about 100 to 200 revolutions, the loss rate decreases following an exponential trend. 
With a revolution half-life of $N_{1/2}=\SI{1481(57)}{}$, this is interpreted as collisional losses with the residual gas in the MR-ToF MS at a pressure of $\SI{3e-8}{\milli\bar}$. 
Using the kinetic radius $r_\text{Kr}=\SI{1.8}{\angstrom}$ \cite{Breck1974} of krypton and $r_{\text{N}_2}=\SI{1.82}{\angstrom}$ \cite{Ismail2015} of the nitrogen molecule, the mean free path $\lambda$ of the ions is estimated to be
\begin{equation}\label{eq:meanfreepath}
    \lambda = \frac{1}{n\,\pi\left(r_\text{Kr}+r_{\text{N}_2}\right)^2},
\end{equation}
where $n$ is the molecular density of the nitrogen gas as given by the ideal gas law for a temperature of $T=\SI{295}{\kelvin}$.
Eq. (\ref{eq:meanfreepath}) leads to $\lambda\approx\SI{3300}{\meter}$, which yields, with a revolution length of $\SI{1.6}{\meter}$, a value of $N_{1/2}\approx\SI{1430}{}$ revolutions, in agreement with the experimental value.

%%%%%%%%%%%%%% RFQcb Characterization %%%%%%%%%%%%%%
\subsection{Radio-frequency quadrupole cooler-buncher}\label{sec:RFQcb}
\begin{figure}[ht]
	\centering
	\includegraphics[width = 0.48\textwidth]{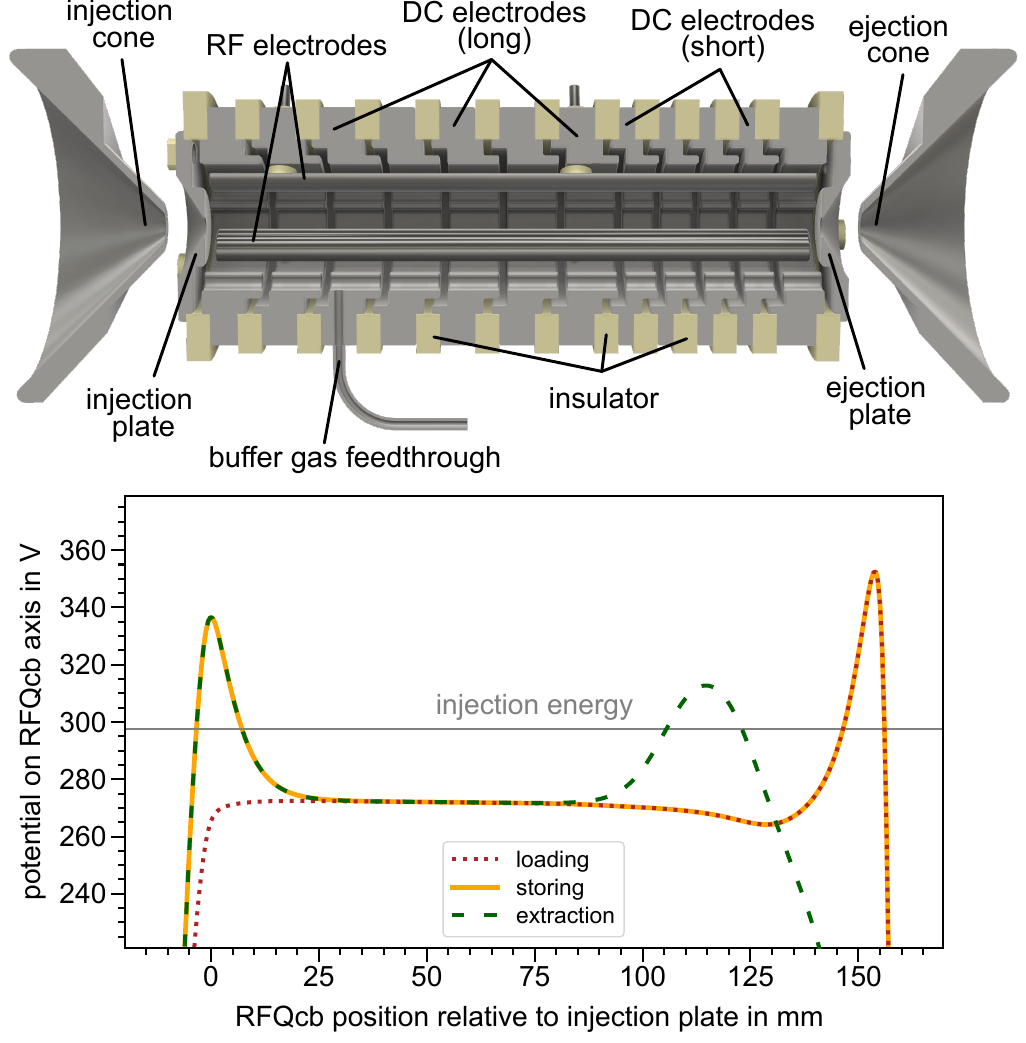}
	\caption{Cross-sectional view of the RFQcb (top) together with the corresponding potential profile on the symmetry axis (bottom). The three operating modes are shown: ion loading (red, dotted), cooling and storage (yellow, solid) and extraction of ion bunches (green, dashed).}
	\label{fig:RFQcb}
\end{figure}

\begin{figure}[ht]
	\centering
	\includegraphics[width = 0.48\textwidth]{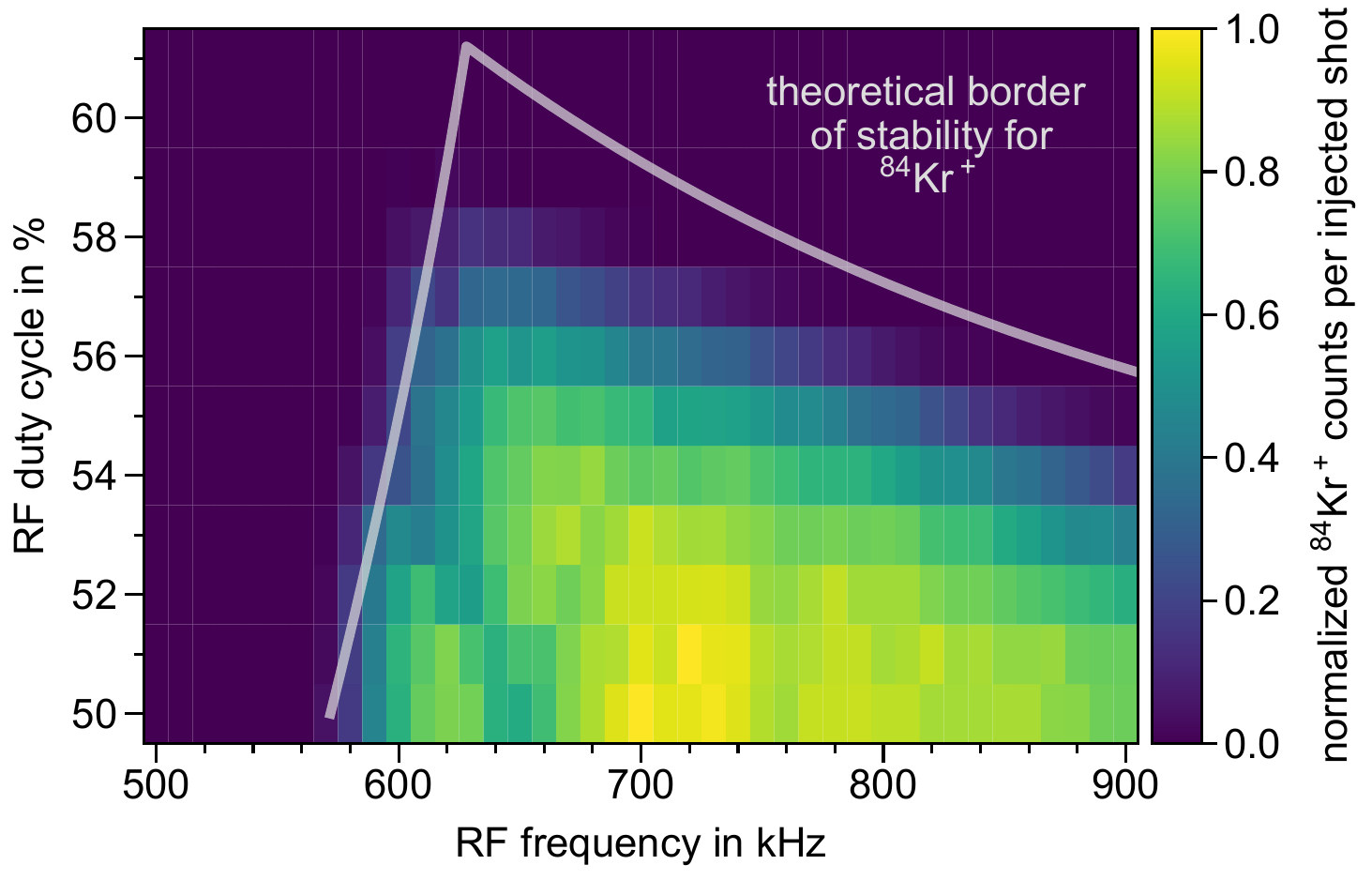}
	\caption{$^{84}$Kr$^+$ beam intensity after accumulation and cooling in the RFQcb as a function of the RF frequency and duty cycle. The boundaries of theoretical stability are indicated for $r_0=\SI{10}{\milli\meter}$.}
	\label{fig:RFQcb_stability}
\end{figure}

\begin{figure}[ht]
	\centering
	\includegraphics[width = 0.48\textwidth]{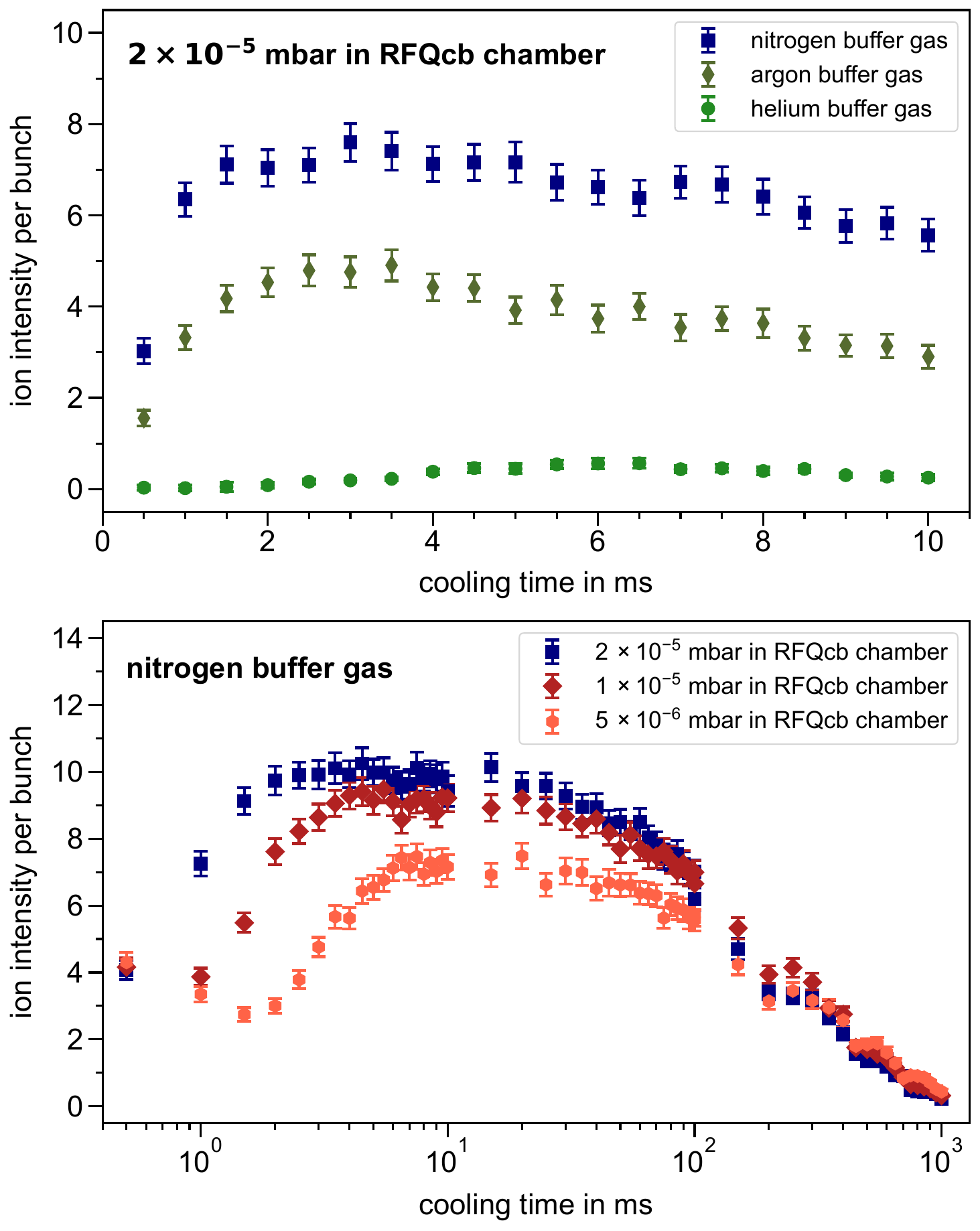}
	\caption{Ion intensity of $^{84}$Kr$^+$ ion bunches released from the RFQcb as a function of the cooling time. Nitrogen, argon and helium set to the same operating pressure are compared (top). Using nitrogen as buffer gas, different operating pressures are compared (bottom). The specified pressures take into account the gas-dependent conversion factors of the pressure gauges.}
	\label{fig:Kr_cooling}
\end{figure}

The RFQcb that is used for the accumulation and cooling of the purified ion beam has been designed within the MIRACLS project \cite{Kanitz2021,Croquette2023} and copies of it were built as part of a collaboration between the Massachusetts Institute of Technology, the University of Greifswald \cite{Giesel2024ASputtering} and the Technical University of Darmstadt. A drawing and the on-axis potential for different operation modes are provided in Fig. \ref{fig:RFQcb}.\\
\indent
Between two endcap electrodes that provide axial confinement, twelve cylindrical DC electrodes define a slightly decreasing potential that leads the ions into a minimum in which they are eventually trapped. To facilitate the injection into and the ejection out of the RFQcb, large cone-shaped electrodes terminate both ends of the assembly. The radial confinement is realized by a radio-frequency (RF) quadrupole field that is generated through a square-wave RF potential applied to four rods located close to the trap axis. 
To increase field penetration, each DC electrode is equipped with wedges between the RF rods  pointing towards the axis. The buffer gas that cools the ions through elastic collisions can be introduced into the trap volume via a gas feedthrough.
\\
\indent
Fig. \ref{fig:RFQcb_stability} shows the stability diagram of the RFQcb for the example of trapped $^{84}$Kr$^+$ ions, cooled with nitrogen buffer gas for $\SI{5}{\milli\second}$. 
It shows at what combinations of RF frequency and duty cycle ion trapping is possible and compares it with the theoretical boundary that is derived from the stability condition for the system (see \cite{Giesel2024ASputtering}).
Further discussion of the stability diagram is provided in the \hyperref[appendix]{Appendix}.
As already addressed in \cite{Giesel2024ASputtering}, the deviation on the high-frequency side may be due to the ratio $R/r_0 = \SI{5}{\milli\meter}/\SI{10}{\milli\meter} = 0.5$ of the RF rods' radius $R$ and their distance from the trap center axis $r_0$ not agreeing with the optimal value for the generation of an ideal quadrupole field in a linear Paul trap of about $R/r_0 = 1.15$ \cite{Dayton1954TheMagnet,Douglas2014MassRods,Lee-Whiting1971Semi-analyticalQuadrupoles}.
% However, the stability region is even somewhat more reduced than shown in \cite{Giesel2024ASputtering}, indicating that the deviation from the ideal RF rod geometry may not be the only limiting factor. 
Also, the DC electrodes introduce higher-order field components that are likely to compromise the stability of the trap.
% Since the RFQcb will not be used for mass separation purposes but only for the accumulation and bunching of the purified beam, optimization of the RFQcb stability region towards the theoretical boundaries is not pursued. 
While the geometry may affect the trapping efficiency, the performance is sufficient to enable the accumulation of more than $10^4$ ions per bunch (see Sec. \ref{sec:OIS_operation}). 
In the present study, the RF field is operated with a duty cycle of $\SI{50}{\percent}$ at frequencies optimized for the respective ion mass.
\\
\indent
The trapping efficiency has been studied with respect to the influence of the buffer gas. 
In particular, besides the commonly used buffer gases helium and argon, nitrogen has also been tested. 
It provides the advantages that it can be pumped efficiently and is readily available in most laboratories.~%
The top plot of Fig. \ref{fig:Kr_cooling} provides a direct comparison of the buffer gases in the case of $^{84}$Kr$^+$ trapping. 
It shows the ion intensity per bunch ejected from the RFQcb for a varying cooling time after the injection of one $^{84}$Kr$^+$ packet.
As determined in a reference measurement beforehand using MagneTOF 2, an average of 37 purified ions was ejected from the MR-TOF towards the RFQcb.~%
After the respective cooling time, the bunched ions are reaccelerated to about $\SI{4}{\kilo\electronvolt}$ and measured with MagneTOF 3. 
In each measurement, the respective buffer-gas flow rate was chosen so that a pressure of $\SI{2e-5}{\milli\bar}$ was reached in the RFQcb vacuum chamber. 
\\
\indent
The largest trapping efficiency is reached in the case of nitrogen, reaching a maximum of $\SI{7.6(2)}{}$ ions for a cooling time of $\SI{3}{\milli\second}$. It suggests a transport efficiency from the MR-ToF MS to MagneTOF 3 of $\approx\SI{20(1)}{\percent}$, including ion deceleration, injection into the RFQcb, trapping, cooling, ejection from the RFQcb, reacceleration and beam focusing through the second quadrupole bender. 
While the trapping efficiency is considerably reduced for argon, ions are barely trapped in the case of helium. To achieve a comparable trapping efficiency in the latter case, higher buffer-gas pressures are needed, with which the specified vacuum requirements cannot be satisfied. 
% \textcolor{red}{Explanation for difference and statement, that helium cannot be used like this}. 
For all three buffer-gas types, the ion intensity rises for short cooling times before it reaches a maximum and steadily decreases again. This is caused by the combined effects of cooling and charge-exchange processes \cite{DELAHAYE2004ChargeExchange,Dilling2001PhD}.
For the cases of nitrogen and argon and the given buffer-gas pressure, the cooling process takes about $\SI{2}{\milli\second}$, indicated by the maximum ion intensity and the ToF data (not shown). 
After the maximum, the rate of ion loss depends on the chosen buffer-gas pressure (bottom, Fig. \ref{fig:Kr_cooling}). 
Since the actual pressure inside the RFQcb cannot be measured, the vacuum chamber pressure is considered for the comparison ("effective pressure"). 
For the tested pressure range, the cooling process is accelerated (decelerated) with increased (decreased) buffer gas pressure. 
At the same time, the impact of the charge exchange is increased (decreased) dominating the trapping efficiency for longer storing times.
The highest trapping efficiency has been reached for an effective pressure of about $\SI{2e-5}{\milli\bar}$.  
However, since the accumulation process may take several $\SI{100}{\milli\second}$, it can be beneficial to reduce the buffer gas pressure below this value to profit from the reduced influence of the charge exchange at longer trapping times. 
It should be noted that these considerations are irrelevant for ion species with negligible charge exchange with the present buffer gases.

%%%%%%%%%%%%%% OIS beamline operation %%%%%%%%%%%%%%
\section{Stacking-mode operation}\label{sec:OIS_operation}
For the intended applications of the POLIS, the chopping and purification process is repeated multiple times until the desired ion intensity of >$10^4$ ions per bunch is accumulated. Considering that single-ion counting loses its validity in this regime, the ion intensity is determined in the following using the integration method according to Eq. \ref{eq:calibration} (see Sec. \ref{sec:detectors}). Analogous to the characterization discussed in \cite{Simke2024EvaluationBunches}, the calibration factor $c_\text{det}=\SI{207(3)}{\text{ions/}\nano\volt\second}$ has been determined for MagneTOF 3 for an operation voltage of $U_\text{det}=\SI{-2150}{\volt}$.

\begin{figure}[ht]
	\centering
	\includegraphics[width = 0.48\textwidth]{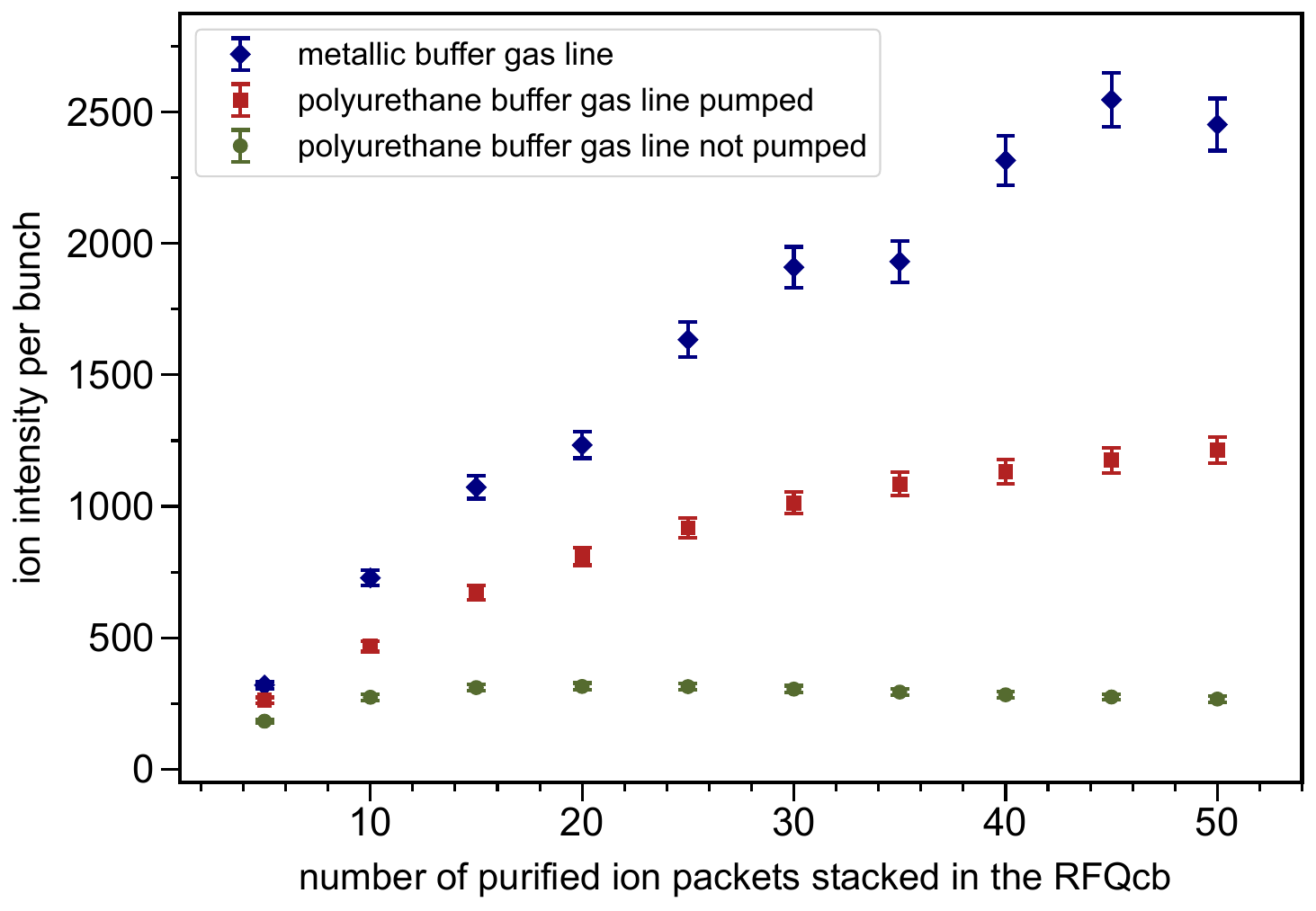}
	\caption{Ion intensity of $^{84}$Kr$^+$ ion bunches released from the RFQcb as a function of the number of ion packets accumulated. Different properties of the buffer gas supply line which lead to different buffer gas purity conditions are compared.}
	\label{fig:stacking}
\end{figure}
\begin{figure*}[ht]
	\centering
	\includegraphics[width = 0.96\textwidth]{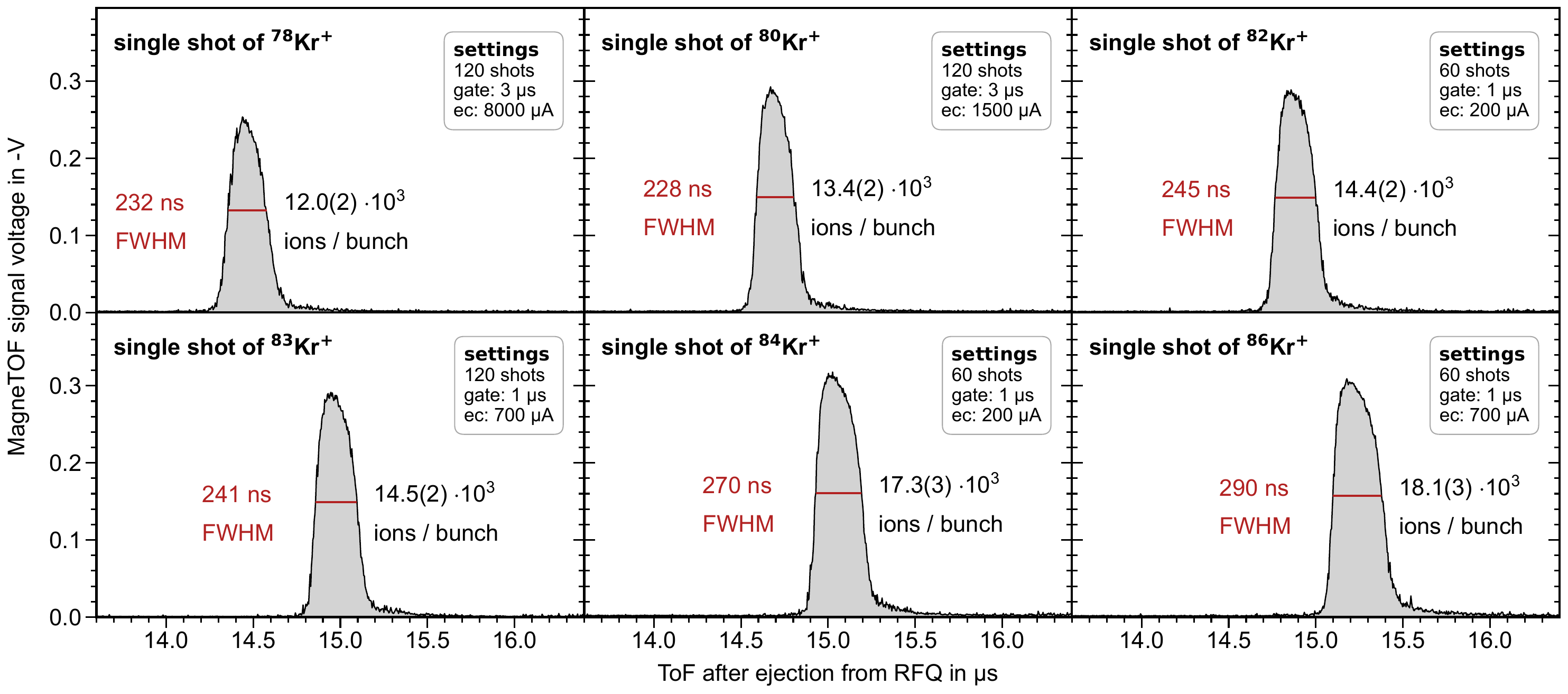}
	\caption{Examples of isotopically pure ion bunches with >$10^4$ ions per bunch for all stable krypton isotopes. The relevant operation settings are given in each plot: the number of shots accumulated, the ion source chopping gate (gate) and the electron emission current (ec).}
	\label{fig:high-intensity}
\end{figure*}

As discussed in the previous section, the achievable ion intensity is affected by charge exchange  with contaminants in the buffer gas or with the buffer gas itself.~%
The extent of this influence is illustrated in Fig. \ref{fig:stacking}, showing the $^{84}$Kr$^+$ intensity per bunch ejected from the RFQcb as a function of the number of purified ion packets stacked in the RFQcb beforehand.~%
The three curves correspond to different conditions of the buffer gas supply line.~%
In the case of the green circles and the red squares, polyurethane tubing with a length of $\SI{1.5}{\meter}$ was used. 
If the experiment is operated for several hours, contaminants accumulate in the buffer gas reservoir through leakage and outgassing, increasing the influence of charge exchange. 
As a consequence, the ion intensity can only be increased by stacking up to 15-20 packets before it saturates and even starts to decrease again (green circles). 
If, on the other hand, the buffer gas reservoir is evacuated and re-filled with high-purity buffer gas right before the measurement, the amount of contaminants can be minimized, reducing the ion losses. 
Accordingly, more shots can be accumulated, and the achievable ion intensity is significantly higher (red squares). 
This experience demonstrates the importance of a dedicated buffer-gas supply system that is leak-tight and minimizes internal outgassing. 

It was therefore exchanged with a supply system fully based on metallic components, which was leak-tested to a value of $<\SI{5e-10}{\milli\bar\liter\per\second}$ (blue diamonds). 
This allows for a further reduction of losses through charge exchange, which ultimately leads to higher intensities.\\
In the measurement previously discussed, the electron emission current of the ion source that defines the extracted ion current has only been set to $\SI{5}{\percent}$ of the maximum value of $\SI{10}{\milli\ampere}$. 
While the number of successively stacked ion packets is limited by charge exchange, increasing the initial ion beam intensity or increasing the duration of the chopping gate applied to the ion source deflector provides an additional approach to obtain higher intensities in the RFQcb. 
This allowed the accumulation of the targeted intensity of >$10^4$ ions per bunch for isotopes with abundances below $\SI{1}{\percent}$. 
Corresponding results are shown for all stable krypton isotopes in Fig. \ref{fig:high-intensity}, for which the chopping-gate duration, electron-emission current and the number of stacked packages have been varied in order to reach comparable bunches.
The corresponding ion intensity, that is determined using the calibration factor mentioned above, as well as the FWHM of the individual signals are indicated in each graph. 
Even for $^{78}$Kr$^+$, which has a natural abundance of only $\SI{0.36}{\percent}$, $\SI{1.20(2)e4}{}$ ions could be accumulated.
For each isotope, the corresponding bunches are extracted with a FWHM smaller than $\SI{300}{\nano\second}$, which is equivalent to about $\SI{30}{\milli\meter}$ for the given transport energy of $\SI{4}{\kilo\electronvolt}$. 
They will therefore be accepted by the $\SI{146}{\milli\meter}$ long Penning trap injection PDT for any targeted energy reduction.
Enriched targets will be considered to achieve similar intensities for isotopes with natural abundances significantly below $\SI{0.1}{\percent}$. 

%%%%%%%%%%%%%% Vacuum estimations %%%%%%%%%%%%%%
\section{Vacuum considerations}\label{sec:vacuum}
Besides its accumulation capabilities, the setup must provide the purified ion bunches at the discussed vacuum level of better than $<\SI{5e-10}{\milli\bar}$.
Starting with a pressure of $\approx\SI{2e-5}{\milli\bar}$ inside the RFQcb chamber, the vacuum level is to be reduced by about five orders of magnitude.
Inevitably, a bake-out of the last transfer beamline section is required, which was applied for $\SI{48}{\hour}$ at $\SI{150}{\celsius}$. 
As a result, a base pressure of $\SI{2e-10}{\milli\bar}$ was reached, measured directly at the junction to the antiproton beamline. 
The same base pressure is reached in the second last vacuum chamber of the transfer beamline.
It provides some margin for the pressure increase due to the injection of the buffer gas in the RFQcb chamber.

\begin{figure}[ht]
	\centering
	\includegraphics[width = 0.48\textwidth]{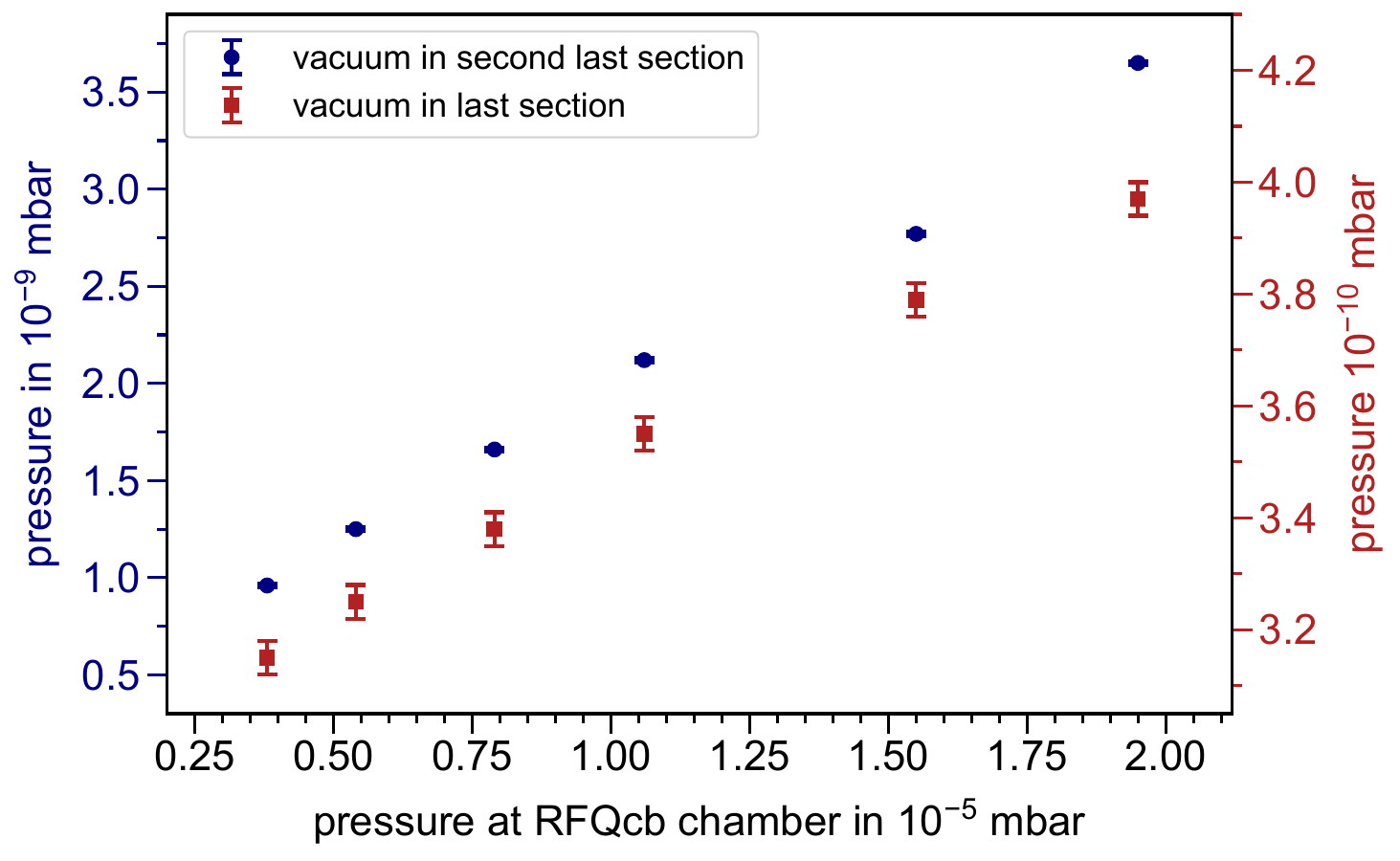}
	\caption{Pressures reached in the last (red squares) and second last (blue circles) differentially pumped vacuum chamber of the POLIS transfer beamline as a function of the pressure in the RFQcb chamber for typical operation values. With no buffer gas injection, the base pressure is at $\SI{2e-10}{\milli\bar}$ for sections.}
	\label{fig:pressure}
\end{figure}

For an experimental test, the opening diameter of the first two iris apertures behind the RFQcb (iris 3 \& 4, compare Fig. \ref{fig:Offline_ion_source}) was reduced such that no transmission losses could be seen at MagneTOF 4. 
Furthermore, iris 5 was set to the same opening diameter as iris 3, since the identical assembly is used at this position. 
For all apertures, this corresponds to an opening diameter of about $\SI{5}{\milli\meter}$.
Varying the buffer-gas pressure (nitrogen) in the RFQcb chamber between $\SI{0.5e-5}{\milli\bar}$ and $\SI{2e-5}{\milli\bar}$, i.e., the typical range of operation, the resulting pressure in the last and second last vacuum chamber of the transfer beamline has been measured (see Fig. \ref{fig:pressure}).
While the pressure rises to a few $\SI{e-9}{\milli\bar}$ in the second last chamber, a value of better than $\SI{4e-10}{\milli\bar}$ can be maintained in the last chamber for all pressures tested.

%%%%%%%%%%%%%% conclusion %%%%%%%%%%%%%%
\section{Conclusion and outlook}\label{sec:conclusion}
An offline ion source (POLIS) beamline, capable of generating cooled and isotopically pure ion bunches, has been developed and commissioned to be used for the PUMA experiment at CERN. 
It provides the isotopes of interest for an experimental program on stable nuclei at the ELENA facility.
The setup includes an electron-impact ionization source for ion generation from gaseous target materials, an MR-TOF MS for beam purification, and an RFQcb for ion accumulation and cooling. 
Despite the broad energy distribution of the extracted ions that originates from the beam chopping, it was demonstrated that ions can be trapped for 1500 revolutions, which allows the separation of any ion of interest from stable multi-species ion beams even in the presence of molecular isobaric contaminants. 
Using the example of stable krypton isotopes, it was demonstrated that the desired intensity of more than $10^4$ ions per bunch can be reached for all isotopes if multiple purified ion packets are stacked in the RFQcb, where they are cooled to a longitudinal FWHM of less than $\SI{300}{\nano\second}$. 
In order to fulfill the strict vacuum requirements at the junction point from the POLIS beamline to the PUMA antiproton beamline, which requires a pressure below $\SI{5e-10}{\milli\bar}$, it was found to be beneficial to use nitrogen as a buffer gas rather than argon or helium for which larger pressures would be needed. 
Successful operation, maintaining this pressure limit, was made possible by the use of adjustable apertures, which significantly limit the conductance of the buffer gas from the RFQcb to the junction to the PUMA antiproton beamline.
It is expected that light ions require helium or hydrogen as buffer gas, which has yet to be tested with regard to the vacuum restrictions set by the PUMA experiment.
Nevertheless, the present measurements show that the system is ready for operation.
It is about to be shipped to CERN, where it will be re-commissioned.

\section*{Acknowledgments}
We wish to express our appreciation to the \mbox{MIRACLS} collaboration for the design and development of the RFQcb.
We are grateful for discussions with the staff of the mechanical workshop at TU Darmstadt and appreciate their efforts in machining a significant part of the in-vacuum components. 
Special thanks also go to D. Neidherr for his significant role in the successful development of the control system.
M.S. would also like to thank M. Bajdak for the in-depth discussions that contributed to an improved stability of the RFQcb. 
The PUMA project is funded by the European Research Council through the ERC grant PUMA-726276. 
C.K., A.O. and M.S. appreciate the support from the Alexander von Humboldt Foundation.

\appendix
\section{Influence of the ion-injection energy}\label{appendix}
\begin{figure}[ht]
    \centering
    \includegraphics[width=0.48\textwidth]{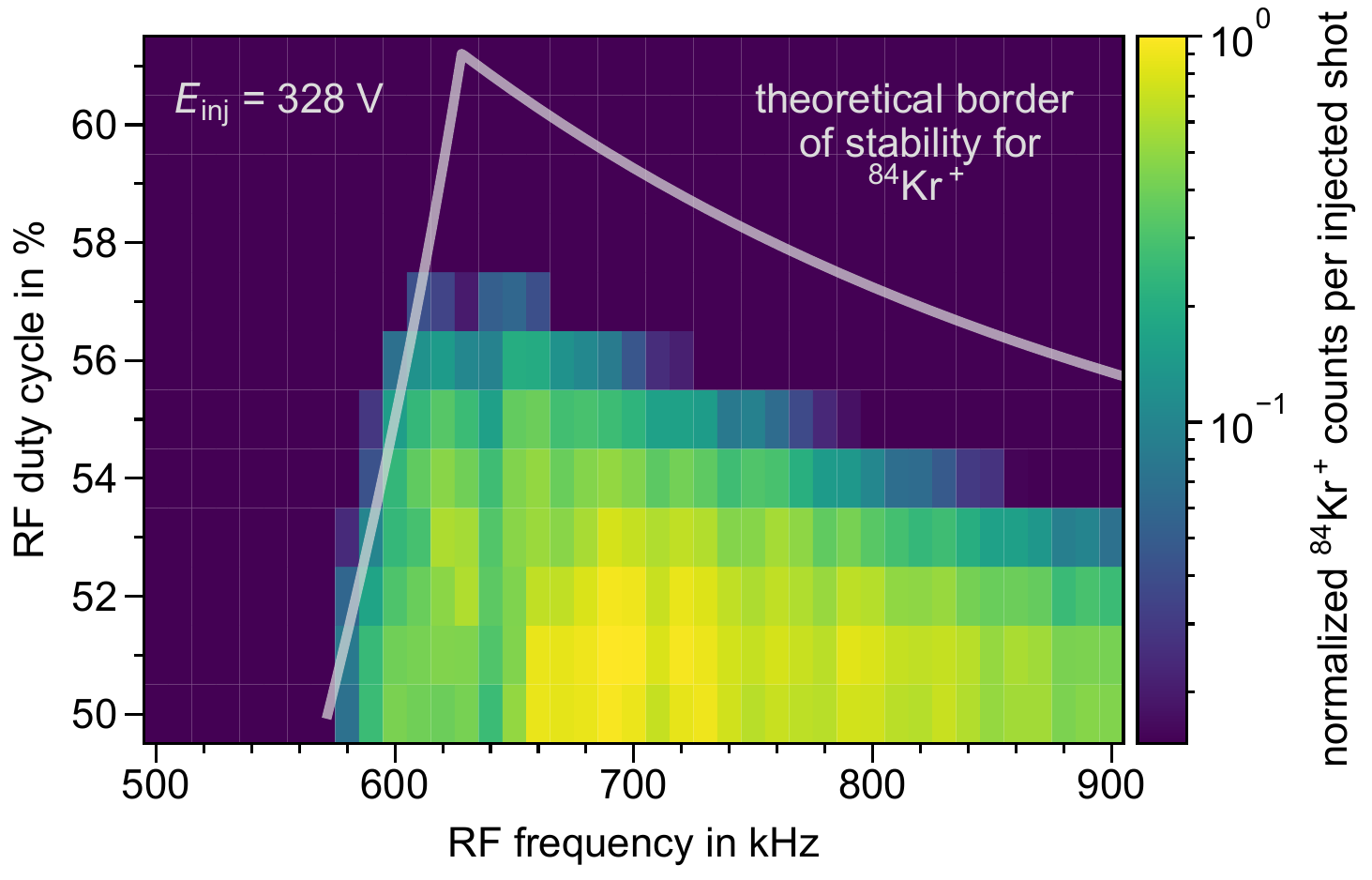}
    \includegraphics[width=0.48\textwidth]{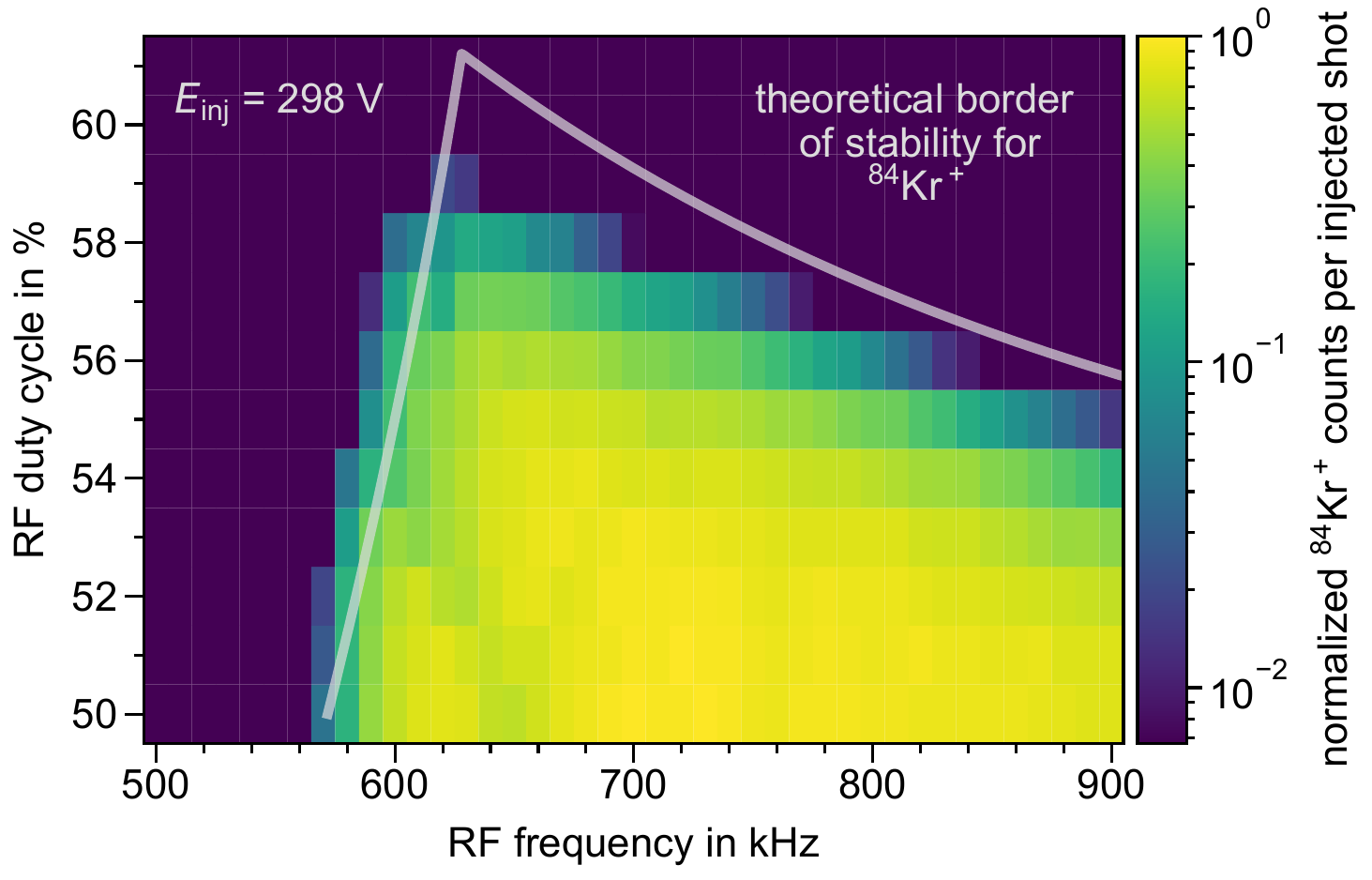}
    \caption{Stability diagrams of $^{84}$Kr$^+$ ions for an ion-injection energy of $\SI{328}{\electronvolt}$ (top) and $\SI{298}{\electronvolt}$ (bottom). To increase the visibility of the low-intensity region, the data is displayed in a logarithmic representation where data points with zero counts are intentionally set to the minimum of the color code.}
    \label{fig:app:stability_new}
\end{figure}

During the characterization of the POLIS beamline, it was found that the RFQcb stability benefits from a reduced ion-injection energy $E_\text{inj}$. 
Maximizing the trapping efficiency for a fixed duty cycle $d$ of $\SI{50}{\percent}$, an initial optimization has led to $E_\text{inj}=\SI{328}{\electronvolt}$. 
A second iteration, however, also taking into account the efficiency for $d>\SI{50}{\percent}$, converged to $E_\text{inj}=\SI{298}{\electronvolt}$.
% Although the optimized RF and DC electrode potentials of both $E_\text{inj}$ are found to be different, the resulting potential profiles on the RFQcb symmetry axis are effectively the same (see Fig. \ref{fig:RFQcb}).
While the maximum of the trapping efficiency for a given combination of $d$ and $\nu$ was found to be comparable for both injection energies tested ($\approx\SI{20}{\percent}$ for $^{84}$Kr$^+$), the average trapping efficiency for all possible values of $d$ and $\nu$ was significantly increased in the case of $E_\text{inj}=\SI{298}{\electronvolt}$. 
A comparison of the respective stability diagrams is provided in Fig. \ref{fig:app:stability_new} and illustrates the increased RFQcb stability for the reduced ion-injection energy. 
% This gets clear comparing both stability diagrams

% \newpage
\bibliographystyle{elsarticle-num} 

\bibliography{references} %mastertheses

\end{document}